\documentclass[preprint,12pt,numbers,square]{elsarticle}

\usepackage{amssymb}
\usepackage[]{amsmath}
\usepackage{mathrsfs}
\usepackage{mathtools} 
\usepackage{enumerate}
\usepackage{caption}
\usepackage[dvipsnames,svgnames,table]{xcolor}
\usepackage{IEEEtrantools}
\usepackage{graphicx}
\usepackage{subcaption}
\usepackage{ragged2e} 
\usepackage{array}
\usepackage{multirow}
\usepackage{verbatim}
\usepackage{soul}
\usepackage{tabularx}
\usepackage{booktabs,siunitx}
\usepackage[T1]{fontenc}
\usepackage{hyperref}
\usepackage{physics}
\usepackage{siunitx}
\usepackage{makecell}
\usepackage{float}
\usepackage{adjustbox}

\AtBeginDocument{\RenewCommandCopy\qty\SI}

\journal{International Journal of Electrical Power and Energy Systems}

\begin{document}

\begin{frontmatter}



\title{Empirical cross-system meta-analysis of long-term transmission grid evolution} 

\author[label1]{Bálint Hartmann}
\affiliation[label1]{organization={Department of Electric Power Engineering, Budapest University of Technology and Economics,},
            addressline={Egry J. u. 18.},
            city={Budapest},
            postcode={H-1111},
            country={Hungary}}
\author[label2,label3]{Michelle T. Cirunay}
\affiliation[label2]{organization={Institute of Technical Physics and Materials Science, HUN-REN Centre for Energy Research},
            addressline={P.O. Box 49},
            city={Budapest},
            postcode={H-1525},
            country={Hungary}}
\affiliation[label3]{organization={Dr. Andrew L. Tan Data Science Institute, De La Salle University},
            addressline={Taft Avenue},
            city={Manila},
            postcode={2401},
            country={Philippines}}

\begin{abstract}
The potential of grid‑side flexibility, the latent ability to reconfigure transmission network topology remains under‑used partly because of the lack of empirical studies on how real‑world grids evolve. We present the first cross‑system meta‑analysis of transmission‑grid evolution by assembling and examining a harmonized annual topology panel for four European systems. Our multi‑decade study covers the Czech Republic, Hungary, the Netherlands, and Slovakia over 50–70 years. For every year and country, we compute a consistent suite of complex network metrics together with multiple 3‑ and 4‑node network motifs. We augment these with an asset‑age layer that reveals how structural features created in distinct historical build‑out phases persist (or fade) through time.

We present evidence that networks were not only expanding but also increasing in complexity. The structural changes of the topology initially decreased the efficiency of the networks, which process later stabilized. The long-term ratio of various network motifs indicate an extended period of the evolution, where an equilibrium between robust and efficient topologies is seen. 
 
 By coupling evidence from longitudinal data with topological characteristics, the study contributes to the understanding of topology control strategies in empirical evolution patterns.
\end{abstract}

\begin{graphicalabstract}
\end{graphicalabstract}

\begin{highlights}
\item First harmonized multi-decade panel of transmission topologies
\item Small-world and motif trajectories reveal different phases of grid evolution
\item Aging and asset renewal linked to topological motifs
\end{highlights}


\end{frontmatter}



\section{Introduction}\label{sec1}
Rapid deployment of renewable energy sources and the intensifying extremes of weather and climate are two important drivers of future infrastructure planning. And while the power industry is traditionally not a risk-taker, system operators and network planners are now expected to look beyond traditional approaches, and exploit the flexibility of the network as well. 
Recent papers on grid-side flexibility echo well to this perspective \cite{Li2018}. While there is considerable literature on optimising generation, storage or consumers to support daily operation, the part potentially played by the grid is untapped, and existing tools are used only on an ad-hoc basis \cite{Nikoobakht2017,policycommonsCongestionMitigation,Numan2023}.

Turning this latent flexibility to a resource used routinely in system operation requires the understanding of how topologies evolve. This is a key research question of the interdisciplinary fields of power engineering and complex network science. And while significant results were generated over the years, evidence base is typically limited to snapshots.
Standout cross-sectional studies analyzing the power system on multiple countries include the work of Rosas-Casals and Corominas-Murtra \cite{RosasCasalsWIT2009}, Pagani and Aiello  \cite{Pagani2013,Pagani2014}, and Espejo et al. \cite{EspejoPHYSICA2018}. However, all these studies are static.

The number of longitudinal studies is also very limited, mostly due to the lack of historic datasets. Buzna et al. \cite{Buzna2009} analyzed the growth of the French 400 kV using a few snapshots of the long evolution between 1960-2000. Hartmann and Sugár \cite{Hartmann2021} assembled a 70-year dataset (1949-2019) of the Hungarian high-voltage network, concluding that drivers of grid evolution are ever-changing, which is reflected in grid vulnerability as well \cite{Hartmann2021b}. Karpachevskiy et al. \cite{Karpachevskiy2021} developed a spatiotemporal database covering the evolution of Moscow’s high-voltage network from its inception in 1913 through the Soviet era to today. Baardman \cite{uuUnderstandingEvolution} used the data of the Hungarian \cite{Hartmann2021} and the Dutch power system (1924-2021) to examine how synthetic networks can replicate their properties. Very recently Hartmann et al. analyzed the structural evolution of an electric distribution network over a 20-year long timespan \cite{Hartmann2025-iz}.

To compensate the low availability of real-world datasets, various evolution models were proposed. The 50-year growth of a high-voltage network consisting of multiple voltage levels was used by Mei et al. \cite{MeiSPRINGER2011} to evaluate small-worldness. Fang et al. \cite{Fang2013} proposed a dynamic evolving model to examine the node degree distributions.

In the present paper, the authors aim to close this gap by performing a cross-system meta analysis of four power systems. To our knowledge, no multi-country, harmonized time series of power system topology analysis was published before. The year-by-year datasets collected and built by us use the same cleaning rules, thus allowing a direct comparison. We compute the same set of graph measures for all studied countries, which lets us detect possible universality in different phases of the evolution. Besides graph measures, we also present a temporal analysis and a motif search, providing insight to the dynamics of topological changes and inherited weaknesses.

The remainder of the paper is organized as follows. Section \ref{sec2} introduces the topological dataset assemble by the authors, and the various metrics used for the comparison of the four countries. Results are presented in \ref{sec3}, and discussed in \ref{sec4}. Finally, conclusions are drawn and possible directions of future research are presented in \ref{sec5}.

\section{Methods and data}\label{sec2}

\subsection{Topological data}\label{sec2.1}

Present paper analyzes topological data of four European countries: Czechia (CZ), Hungary (HU), The Netherlands (NL) and Slovakia (SK) (see Table \ref{table:countries}). The historical datasets were compiled by the first author from heterogeneous sources—including handwritten notes, anniversary volumes, statistical yearbooks, cartographic materials, and expert consultations. A rigorous pre-processing and standardization was carried our to ensure cross‑source. In the resulting database \cite{Hartmann2024TopologicalData} a node is created for the year in which a substation is first operational, and an edge is created for the year a transmission line enters service. Infrastructure elements are removed from the dataset in the calendar year of their decommissioning.

While the yearly data for the countries cover slightly different periods, they all include the period during which high-voltage transmission systems (above 220 kV) were established. Analyzed periods for CZ, HU, NL and SK were 1951-2012, 1967-2019, 1970-2024 and 1961-2023, respectively. (Note, that before 1993, the split of Czechoslovakia, the CZ and SK systems were developed jointly. The transmission lines connecting the two federal states were not considered in this study, but are included in our database \cite{Hartmann2024TopologicalData} under the Czechoslovakian data.) Additionally, as historical data for sub-transmission levels is available for NL (110 and 150 kV) and Hungary (120 kV), the analysis was extended to these larger networks as well.

\begin{table}[H]
\centering
\begin{adjustbox}{width = 0.8\textwidth}
\begin{tabular}{|c|c|c|}
\hline
\textbf{HV Networks}      & \textbf{Period covered} & \textbf{Number of years} \\ \hline
CZ     & 1951 - 2012    & 62              \\ 
HU     & 1949 - 2019      & 71              \\ 
NL & 1931 - 2024    & 94              \\ 
SL    & 1959 - 2023    & 65              \\ \hline
\end{tabular}
\end{adjustbox}
\caption{\label{tab:datasets} High-voltage networks and time periods in consideration.}
\end{table}

Using this dataset, graph representations were created for each year, with the nodes being generators, transformers and substations and the edges being transmission lines. Double-circuit lines were treated as single connections. These graphs were used to calculate the various metrics, presented in \ref{sec2.2}. Besides these metrics, we searched for simple motifs such as loops and stars. Loops are crucial structures in transmission grids, as they provide redundancy and this flexibility in operation. 3-loops are important contributors to basin-stability \cite{Schultz2014}, and \cite{Dey2017} reported 4-loops are rare in weaker power systems, thus motif concentrations can be used as a metric of fragility. On the other hand, star topologies provide cheap connections, but create vulnerable nodes for targeted attacks. It was therefore expected, that the share of star motifs would decrease over time, and more loops would appear.

\begin{table}[h]
    \centering
    \resizebox{\columnwidth}{!}{
    \begin{tabular}{|l|r|r|r|r|}
        \hline
        \textbf{Country} & \textbf{Size [km$^2$]} & \textbf{Population} & \textbf{El. cons. [TWh]} & \textbf{Res. el. cons. [TWh]} \\
        \hline
        CZ & 78,867 & 10,837,890 & 62.077 & 15.074 \\
        HU & 93,028 & 9,855,745 & 43.186 & 11.677 \\
        NL & 41,543 & 17,772,378 & 111.757 & 22.754 \\
        SK & 49,035 & 5,563,649 & 26.372 & 5.891 \\
        \hline
    \end{tabular}
    }
    \caption{Illustrative comparison of the analyzed countries. Data include 2022 and 2023 facts and estimates.}
    \label{table:countries}
\end{table}

\subsection{Complex network metrics}\label{sec2.2}

To quantify the changes over the years, we employ multiple complementary complex network metrics to assess the evolving structure and performance of the networks.

The link density $D$ measures the ratio of existing edges relative to the maximum possible number of edges in a fully connected network:

\begin{equation}
D = \frac{2E}{N(N-1)}
\label{eqn:density}
\end{equation}

\noindent where $N$ is the number of nodes and $E$ is the number of edges.

The average node degree for an undirected graph is simply the average number of edges per node in the graph:

\begin{equation}
\left< k \right> = \frac{2E}{N}
\label{eqn:ave_degree}
\end{equation}

\noindent where the factor of 2 arises from each edge contributing to the degree of two distinct vertices.

The average path length $L$ is defined as the average number of steps along the shortest paths between all pairs of nodes:

\begin{equation}
L = \frac{1}{N(N-1)} \sum_{i \neq j} d(i,j)
\label{eqn:spl}
\end{equation}

\noindent where $d(i,j)$ is the shortest path between nodes $i$ and $j$.

Another metric to characterize the network is its diameter $d$ which is the longest among all calculated shortest paths in the network. As this quantifies the distance between the two most distant nodes in the network, the diameter of the graph provides a hint on the overall connectivity i.e. a smaller diameter implies better connectivity across the network.

The clustering coefficient $C$ measures the tendency of a node’s neighbors to be connected and is given by:

\begin{equation}
C = \frac{1}{N} \sum_{i} \frac{2E_i}{k_i(k_i - 1)}
\label{eqn:clustering}
\end{equation}

\noindent where $E_i$ is the number of edges between the neighbors of node $i$, and $k_i$ is its degree.

To detect sections of the networks that are highly connected locally but have sparse connections to other clusters, we employ the modularity metric $Q$ \cite{Newman2006-bw} given by

\begin{equation}
Q = \frac{1}{2E} \sum_{ij} \left( A_{ij} - \gamma \frac{k_i k_j}{2E} \right) \delta(c_i,c_j)
\label{eqn:modularity}
\end{equation}

\noindent where $E$ is the total number of edges in the network, $A_{ij}$ is an element of the adjacency matrix (1 if there is an edge between nodes $i$ and $j$), $k_i$ is the degree of node $i$, $c_i$ is the community where node $i$ belongs to, and $\delta(c_i, c_j)$ is the Kronecker delta function whose value is 1 if nodes $i$ and $j$ belong to the same community, and 0 otherwise. If $Q > 0$ indicates good community structure; $Q \approx 0$ no community structure is detected; and if $Q < 0$, although rare, indicates the network is less modular than random.

To provide some sort of measure of network performance over the years, as more lines are added, we also consider the global efficiency $\eta$ which is a measure of efficient power flows in the network \cite{latora2001}. It is defined as:

\begin{equation}
\eta = \frac{1}{N(N-1)} \sum_{i \neq j} \frac{1}{d(i,j)}
\label{eqn:efficiency}
\end{equation}

\noindent High-$\eta$ means that most nodes can reach others easily, even in large networks.

The Watts-Strogatz small-world coefficient $\sigma$, proposed by Fronczak et al. \cite{FronczakPRE2004}, combines clustering and path length to characterize small-world behavior,

\begin{equation}
\sigma = \frac{C/C_r}{L/L_r}
\label{eqn:sigma}
\end{equation}

\noindent where the $C_r$ and $L_r$ are the corresponding clustering coefficient and average shortest path length for random graphs \cite{FronczakPRE2004}, given by:

\begin{equation}
C_r = \frac{\left< k \right>}{N}
\label{eqn:clustering_random}
\end{equation}

\begin{equation}
L_r = \frac{\ln(N) - 0.5772}{\ln \left< k \right>} + 0.5
\label{eqn:spl_random}
\end{equation}

\noindent A network is considered small-world if $\sigma > 1$, meaning it has significantly higher clustering than a random network while maintaining a comparable path length.

An alternative small-world measure, proposed by Telesford et al. \cite{TelesfordBRAIN2011}, addresses the limitations of $\sigma$ by comparing the clustering coefficient to that of a lattice (regular) network. This metric, $\omega$, ranges between -1 and 1:

\begin{equation}
\omega = \frac{L_r}{L} - \frac{C}{C_{\text{lattice}}}
\label{eqn:omega}
\end{equation}

\noindent Values of $\omega \approx 0$ indicate small-world characteristics ($L \approx L_r$, $C  \approx C_{lattice}$). When $\omega > 0$ the network is more random-like ($L \approx L_r$, $C <  C_{lattice}$), while $\omega < 0$ indicates a more regular, lattice-like topology ($L > L_r$, $C  \approx C_{lattice}$)

\section{Results}\label{sec3}
In the following, we present our results, organized in three subsections. Graph measures are shown in Section \ref{sec3.1}, temporal analysis in Section \ref{sec3.2}, and network motifs in Section \ref{sec3.3}

\subsection{Complex network metrics}\label{sec3.1}

\begin{figure}[H]
     \centering    \includegraphics[width=\columnwidth ]{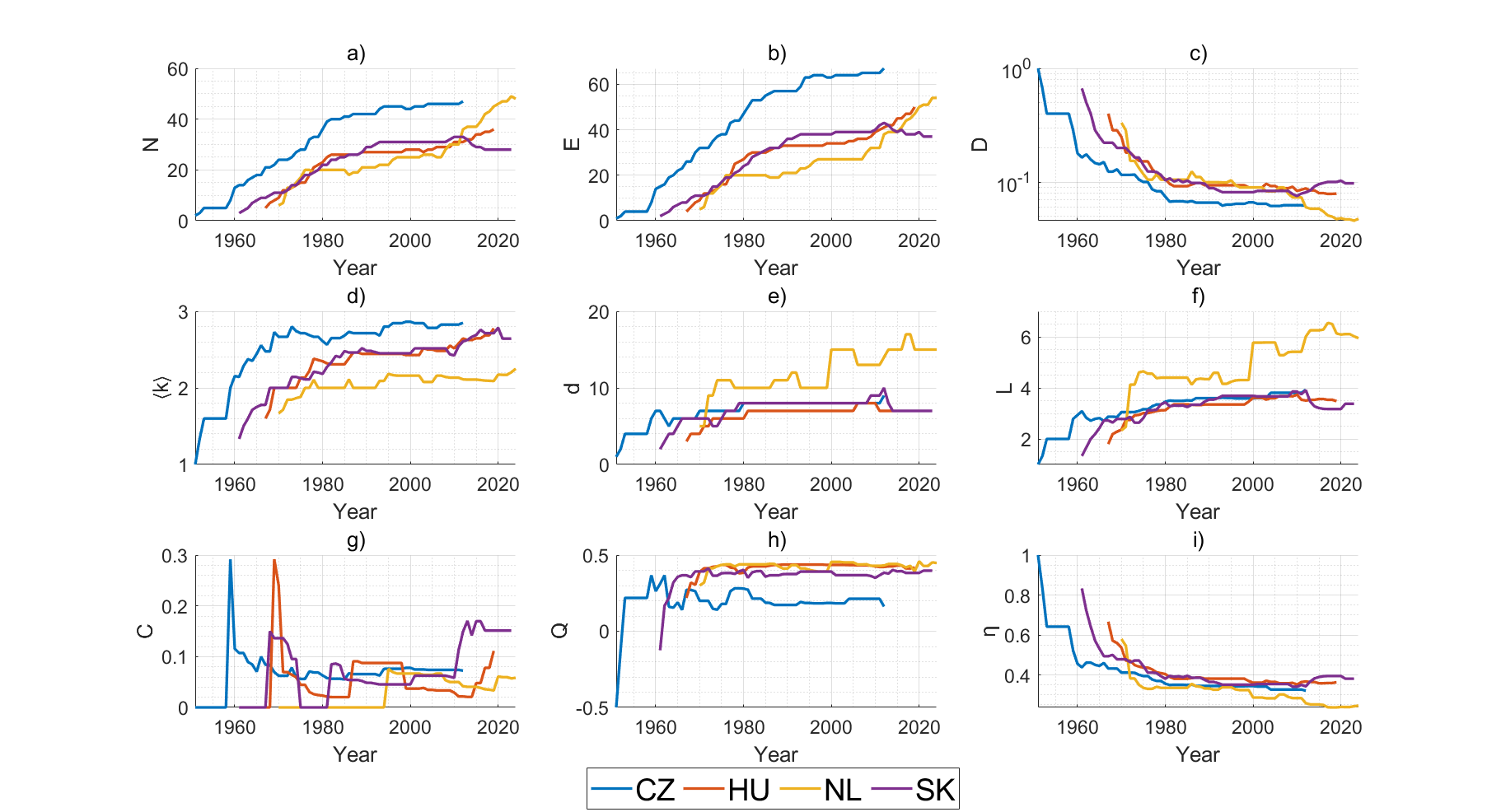}
     \caption{Selected characteristics of the four transmission systems. $N$ and $E$ denote the number of nodes and edges.$D$ is the link density, $\langle k\rangle$ is the average degree, $d$ is the diameter, $L$ is the average path length, $C$ is the clustering coefficient, $Q$ is the modularity quotient, $\eta$ is the global efficiency of the network}\label{fig:CNA_transmission_alt}
 \end{figure}

\begin{figure}[H]
     \centering    \includegraphics[width=\columnwidth ]{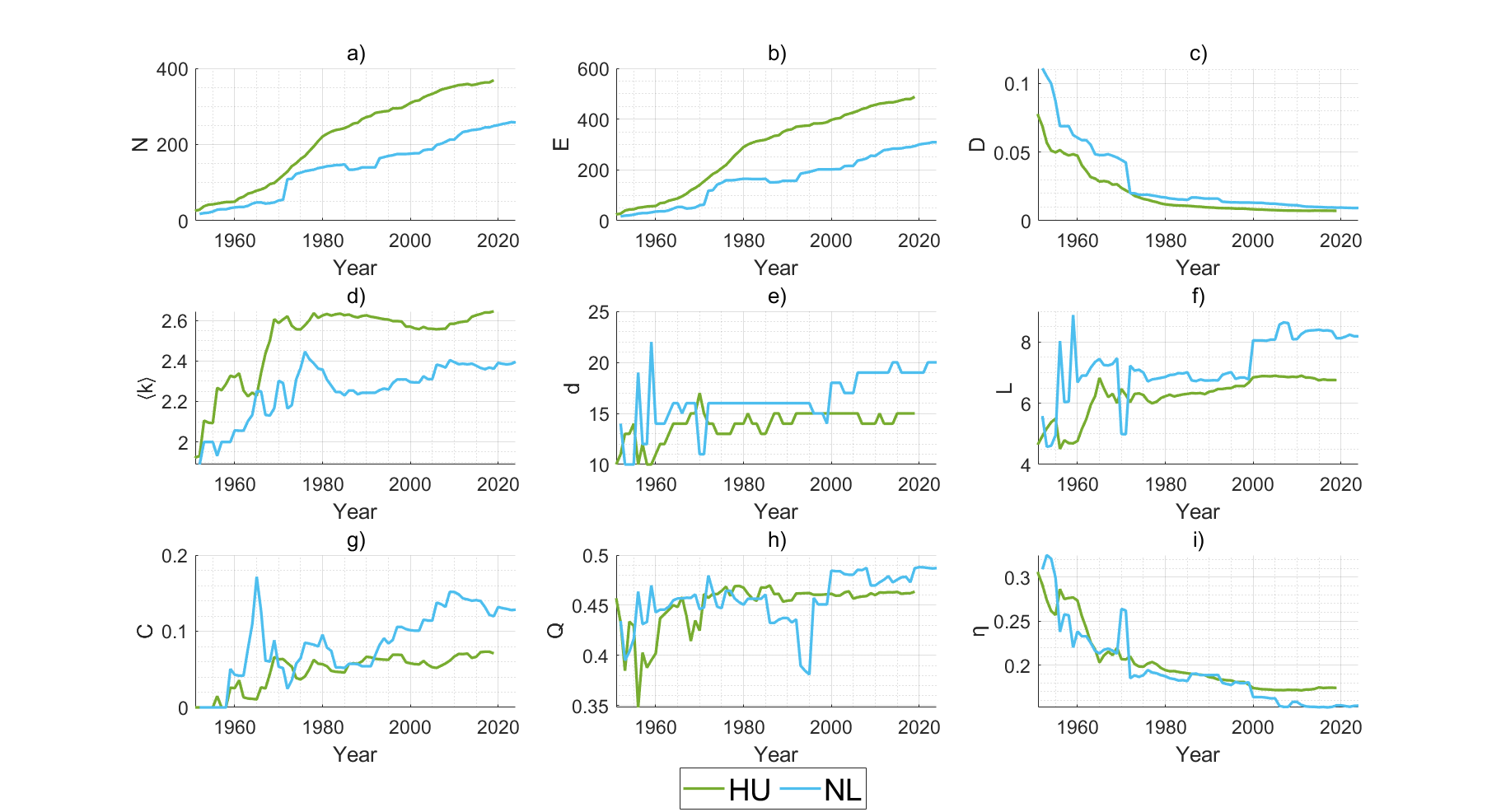}
     \caption{Selected characteristics of the HU and NL high-voltage systems, including transmission and sub-transmission elements. Notations are the same as for Figure \ref{fig:CNA_transmission_alt}.}\label{fig:CNA_subtransmission}
 \end{figure}

For the transmission grids of the four countries in Figure~\ref{fig:CNA_transmission_alt} and similarly for the complete high-voltage networks of HU and NL shown in Figure~\ref{fig:CNA_subtransmission}, we generally observe an increasing trend for most metrics being considered. This signifies an increase in both expanse (increase in the number of nodes $N$, network diameter $d$, average path length $L$, and decrease in link density $D$) and complexity (number of edges $E$, average degree $\left< k \right>$, clustering coefficient $C$, modularity $Q$)of the network. Apart from the expansion, the networks also densified within their initial territory as indicated by the saturation in the trends. Finally, as a result of this growth, we observe a decrease in the global efficiency of the networks over the years.

The increase in topological distance $d(i,j)$ (which in turn decreased the global efficiency) is in many cases the results of increasing consumption density, which necessitated more connections between transmission and sub-transmission levels. This led to the segmentation of originally long lines, increasing the number of nodes (Figures~\ref{fig:CNA_subtransmission} and \ref{fig:CNA_transmission_alt}). Additionally, these expansions are frequently radial (indicated by an increase in average degree shown in Figure~\ref{fig:CNA_transmission_alt}(d)), with single lines feeding outward, which tends to lengthen the shortest paths across the network (Figure~\ref{fig:CNA_transmission_alt}(f)). In some cases, aging infrastructure or underinvestment means that long-range transmission capacity does not keep pace with demand, further increasing effective distances. As a result, lower global efficiency can make the system slower to reroute power after local failures, complicate voltage control and lead to slightly higher energy losses due to more steps in transmission. However, this is not always negative; sometimes it reflects a deliberate design choice to add more local connections for redundancy, enhancing resilience against single line failures.

However, similar to the other metrics, we can also observe a saturation in the decrease of global efficiency. In Figure~\ref{fig:CNA_transmission_alt}(i) the global efficiency is steadily declining which likely indicates outward expansion with more local or radial feeders rather than new cross-network high-voltage ties. The rising trend for the clustering coefficients suggests the grid is becoming more locally clustered, remaining robust in local areas but more congested to transmit power across the entire network.

\begin{figure}[H]
     \centering    \includegraphics[width=\columnwidth ]{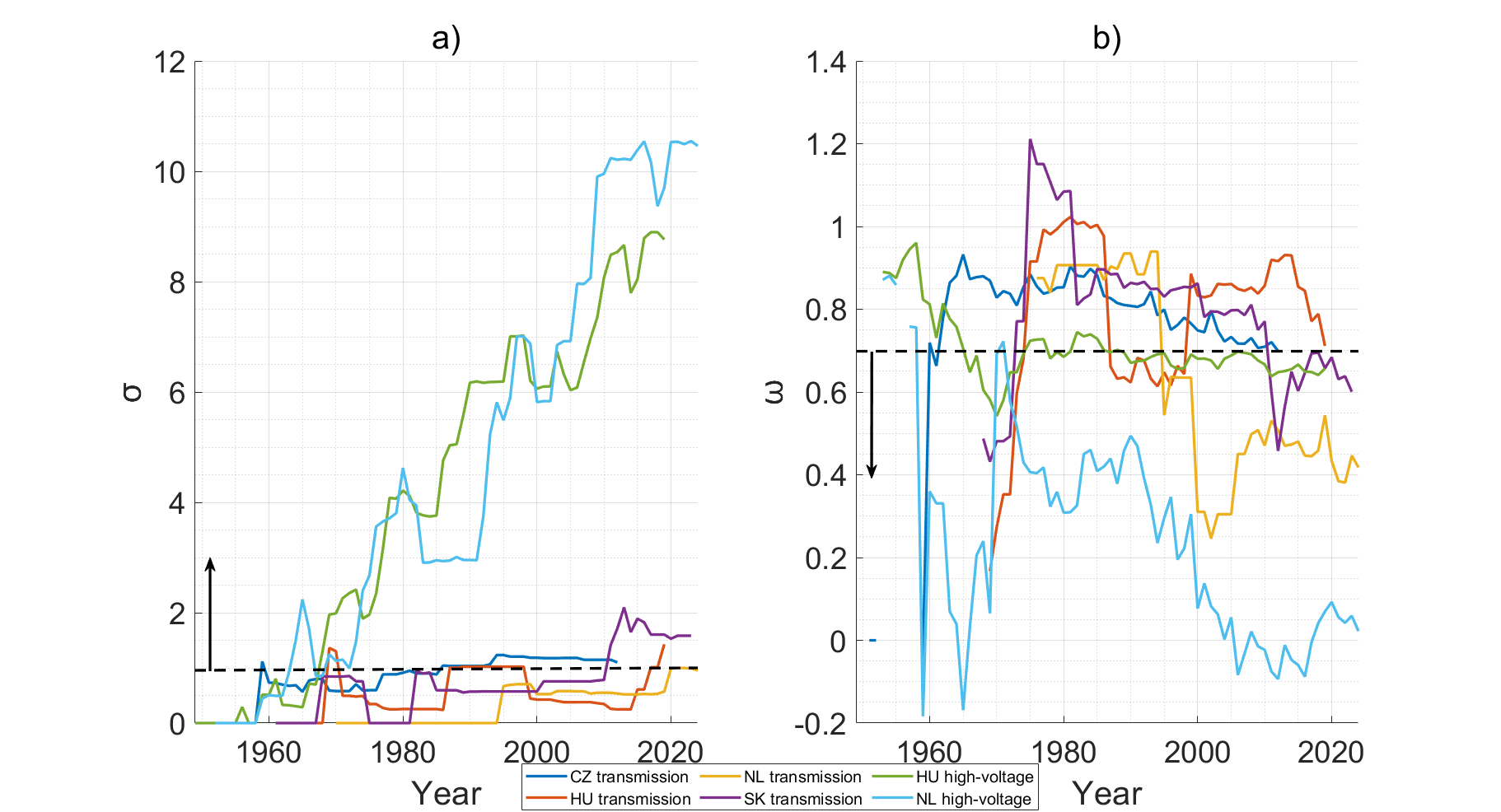}
     \caption{Small-worldness of the four transmission networks. $\sigma$ and $\omega$ are small-world metrics according to Fronczak et al.\cite{Fron} and Telesford et al. \cite{TelesfordBRAIN2011}}\label{fig:SW}
 \end{figure}

Figure~\ref{fig:SW} shows the evolution of the small-world metrics $\sigma$ and $\omega$. In general, we observe that the trends for the transmission networks of the four datasets fluctuate in the non-small-world region for both metrics $\sigma$ and $\omega$. For the case of $\omega$, we treat a looser boundary value of $0.7$ (instead of Telesford's small-world range between $0.5 < \omega < 0.5$) to account for the sparse nature of power grids, similarly as in \cite{Hartmann2021}. To reiterate, a graph is commonly classified as small-world if $\sigma>1$. On the other hand, $\omega$ allows us to have a picture of where our network structure lies: values that are close to zero indicate small-world characteristics, while values closer to -1 suggest a lattice-like structure and values closer to 1 indicate a random graph structure.

In Figure~\ref{fig:SW}(a), we can observe that the transmission networks for all the datasets exhibit $\sigma < 1$, which indicates their non-smallworld (low clustering, long path length) characteristics. This makes sense because transmission networks are built for cost efficiency and safe power delivery where power is routed over long distances through hubs and powerlines. In contrast, the complete high-voltage networks of NL and HU show an increase in smallworldness over the years. As these high-voltage networks include sub-transmission lines, which bridge long-distance transmission and local distribution, and lines that are responsible for inter-regional transfer and tend to form a robust, efficient backbone. Given these two components, this causes the HV networks to be locally clustered for fault tolerance and to have long-distance transmission, characteristics of a small-world network.

In Figure~\ref{fig:SW}(b) we show the evolution of the networks' smallworldness based on Telesford's metric $\omega$. Here, the transmission networks exhibit a decreasing behavior, but not necessarily within the smallworld regime, until the most recent years. For the most part, they exhibit $\omega \geq 0.7$ which indicates that the networks are more random-like in structure characterized by short path lengths that can result in higher global efficiency. On the other hand, the complete high-voltave networks of HU and NL show a rapid initial drop from the non-smallworld regime to being small-world in the earlier years. While NL proceeded to becoming more small-world (or possibly later on, lattice-like since the trend is decreasing), HU have remained stable at the boundary of $\omega \approx 0.7$, and have remained there from 1970s to 2020. Note that $\omega$ is a difference between normalized values of clustering and path lengths.

 \begin{figure}[H]
     \centering    \includegraphics[width=\columnwidth ]{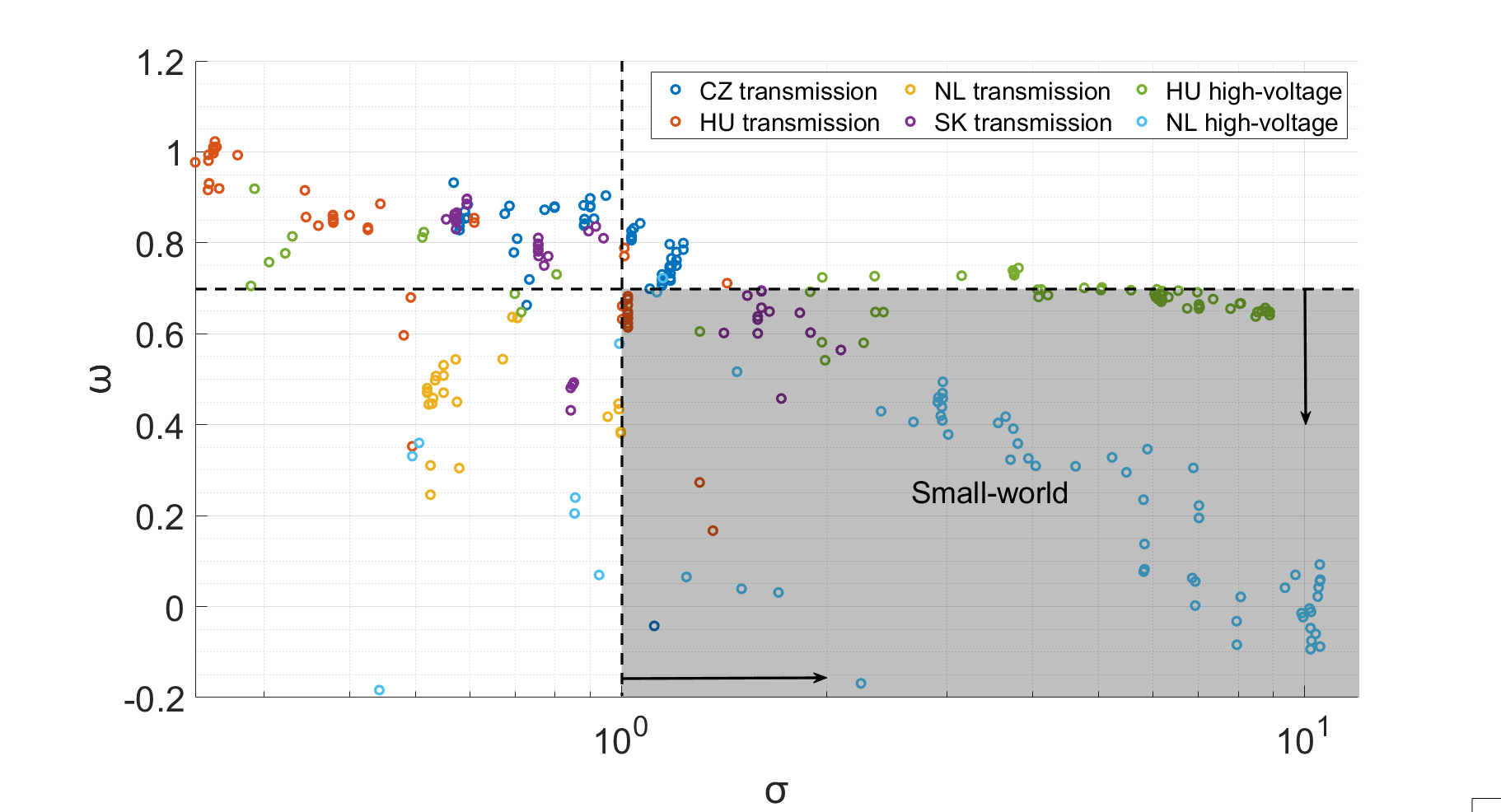}
     \caption{Small-worldness metrics plotted against each other.}\label{fig:SW2}
 \end{figure}

In Figure~\ref{fig:SW2} we show the different metrics of small-world against each other. The resulting plot also visually confirms that if definitions are taken rigorously, the NL high-voltage network is the only one that can be considered as small-world, while the HU high-voltage network is on the edge. These findings align with the results of \cite{Hartmann2021}, implying that small-worldness in power grids is more prominent if multiple voltage levels are considered.

\begin{figure}[H]
     \centering    \includegraphics[width=\columnwidth ]{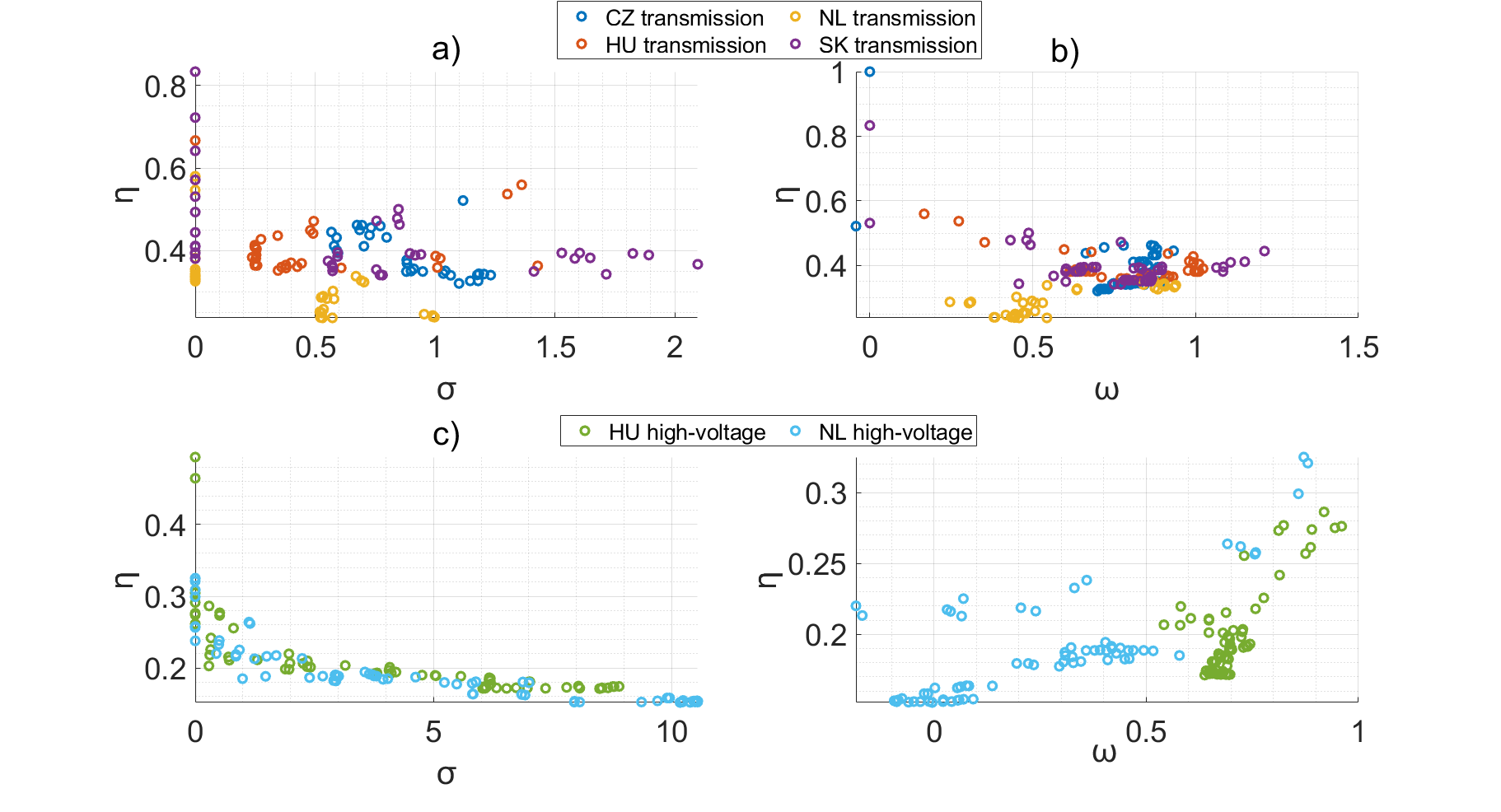}
     \caption{Efficiency plotted against the two small-world metrics}\label{fig:Efficiency}
 \end{figure}

Although the small-worldness of a network captures a balance between high local clustering and short average path lengths (relative to a comparable random network), it does not always translate into higher global efficiency. Figure~\ref{fig:Efficiency}(a)-(b) show that increases in small-worldness over time did not necessarily correspond to increases in global efficiency. This may be because small-worldness is driven by the ratio of normalized clustering to normalized path length, which means it is sensitive to relative changes rather than absolute distances across the network. In contrast, global efficiency directly measures the inverse of shortest path lengths over all pairs, providing a more explicit indication of how efficiently power can propagate through the grid.
In practice, this means that even if the grid evolves to become more "small-world" by improving local redundancy or clustering, it might simultaneously experience growth that increases the overall topological distances (i.e. many new peripheral connections), thereby lowering global efficiency. Thus, while small-world features can indicate improved robustness and fault tolerance at the local level, they do not automatically ensure that the grid becomes more efficient in moving power over long distances.

Despite observable structural changes in the network over time (Figure~\ref{fig:CNA_transmission_alt} and variations in small-worldness, the global efficiency of the power grids studied generally remained steady as shown in Figure~\ref{fig:Efficiency}. This stability suggests that while the grids evolved, adding new nodes and edges and possibly enhancing local clustering, these changes did not significantly impact the overall ability of the network to transfer power efficiently on a global scale. This means that the expansions and modifications of the grid were managed in a way that preserved essential topological performance, maintaining short effective path lengths on average even as the network grew. This highlights a characteristic resilience of the grid’s design: accommodating growth and local redundancy without compromising the fundamental efficiency needed for reliable large-scale operation.

 \begin{figure}[H]
     \centering    \includegraphics[width=\columnwidth ]{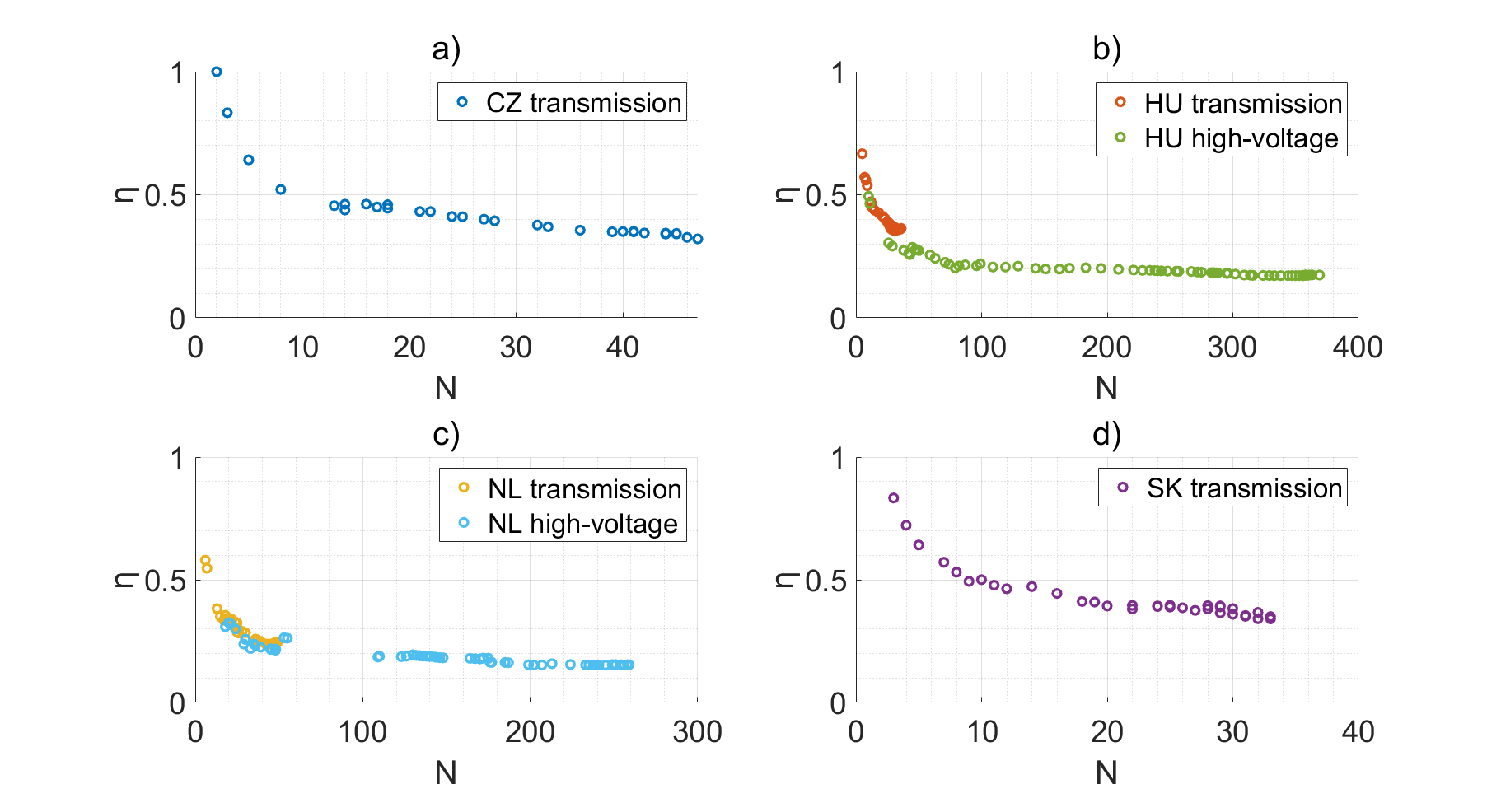}
     \caption{Efficiency as a function of node number}\label{fig:NvsEff}
 \end{figure}

In Figure~\ref{fig:NvsEff} we observe how the global network efficiency changes as more lines of different voltage levels are constructed in the networks. We found that similar to the behavior in Figures~\ref{fig:CNA_transmission_alt}(i) and~\ref{fig:CNA_subtransmission}(i), the network efficiency drops but eventually stabilizes. As previously discussed, this is due to the fact that the latter network modifications is in the form of the addition of substations (nodes) increasing the degree and local clustering, instead of creating long-range powerlines. In the case of HU (Figure~\ref{fig:NvsEff}(b)) and NL (Figure~\ref{fig:NvsEff}(c)), we observe that transmission networks have higher global efficiency than HV networks. This is due to a few reasons such as (i) transmission networks have relatively few nodes and high node degree, creating a strong backbone but low geographical coverage; and (ii) HV networks include a lot more dead-ends as they are to bring electricity to boundary zones as well

\subsection{Temporal analysis}\label{sec3.2}
\begin{figure}[H]
     \centering    \includegraphics[width=\columnwidth ]{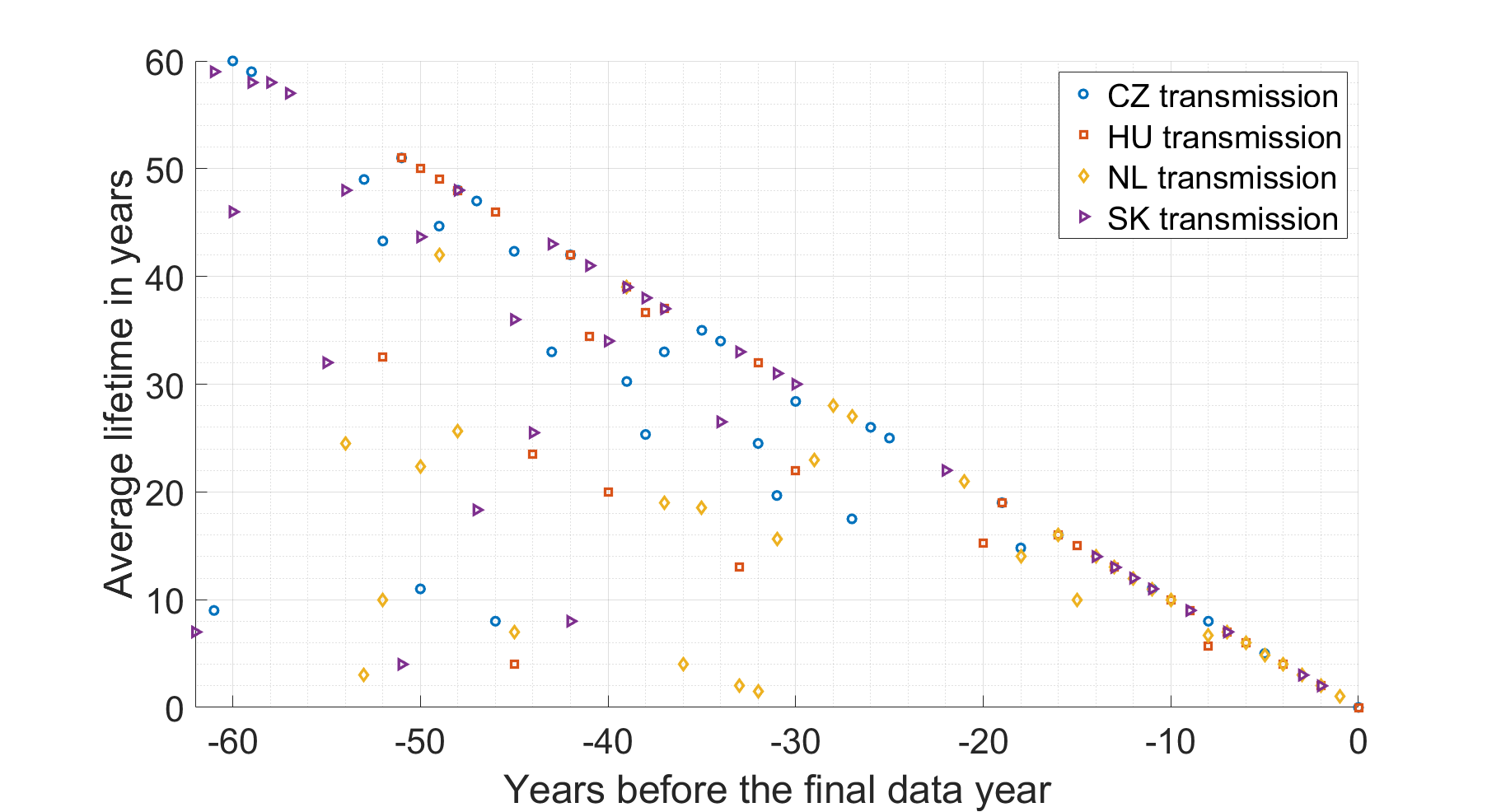}
     \caption{Average lifetime of the lines constructed in given years.}\label{fig:Lifetime}
 \end{figure}

To compare the temporal variations of the topologies of the four countries, we aligned the periods of evolution to each other. First we consider the average lifetime of transmission lines, where the lifetime is defined as the number of years between commissioning and the first major change. This change is typically a topological change, but it may reflect to upgrade and the increase of nominal voltage as well. Figure~\ref{fig:Lifetime} shows that besides minor differences, the evolution of the topology is very similar in the four countries. Typically, a topological formation is not expected to change in the first approx. 25 years of operation, which is in line with the usual approaches of transmission expansion planning. It also highlights that grid flexibility from transmission system could be likely provided through transmission topology control rather than foresight in planning.

\begin{figure}[H]
     \centering    \includegraphics[width=\columnwidth ]{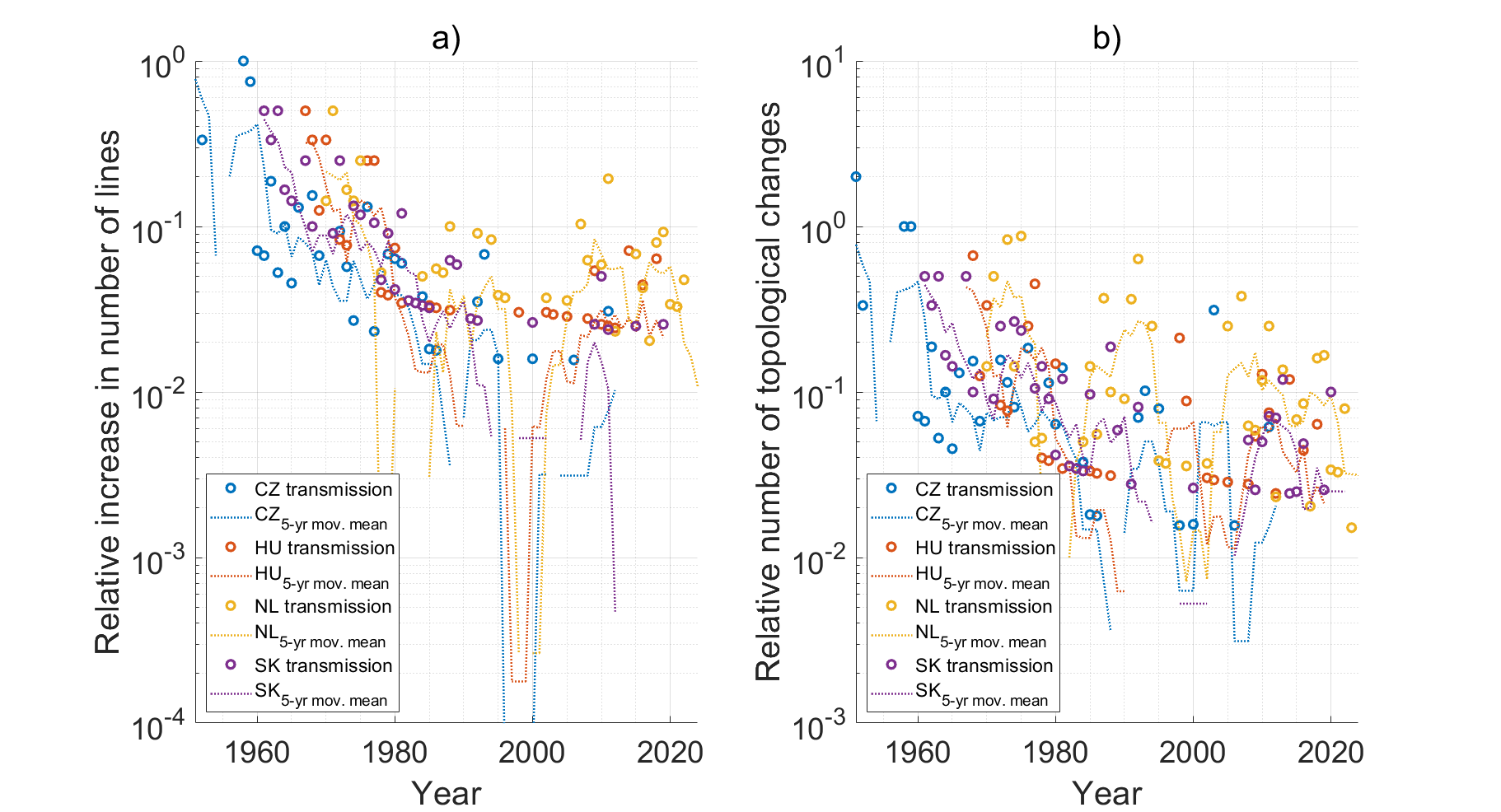}
     \caption{Temporal characteristics of the evolvement the four transmission systems. The number of new power lines and topological changes are normalized to the number of lines in operation in a given year. Moving average is calculated to 5 years.}\label{fig:Temporal}
 \end{figure}

We also analyzed how frequent these topological changes were during the multi-decade dataset; Figure~\ref{fig:Temporal} presents the results on this. For both subfigures, we applied a 5-year moving average to indicate the trends. As Figure~\ref{fig:Temporal}a) shows, the relative number of commissioned transmission lines shows a steady decrease from the early 1980s. By this time, all of the analyzed countries have constructed the backbone of their transmission system, and covered the majority of the countries' area. (The first >220 kV lines went online in 1960, 1977, 1969, and 1968 for CZ, HU, NL and SK, respectively.) It can be seen that the NL shows more active development in the 1990s; this was the era, when the first 380 kV loop was created, thus effectively creating a multi-layered network of >220 kV lines. This difference is also captured by Figure~\ref{fig:Temporal}b), as the relative number of topological changes reaches a local peak at the same time for NL. Another common trend is an increase in changes in the 2010s, which can be already connected to the transitioning to more distributed energy sources, and the phasing out of large fossil-fueled power plants, constructed after World War II - a process leaving many substations obsolete.

\subsection{Motif search}\label{sec3.3}

\begin{figure}[H]
     \centering    \includegraphics[width=\columnwidth ]{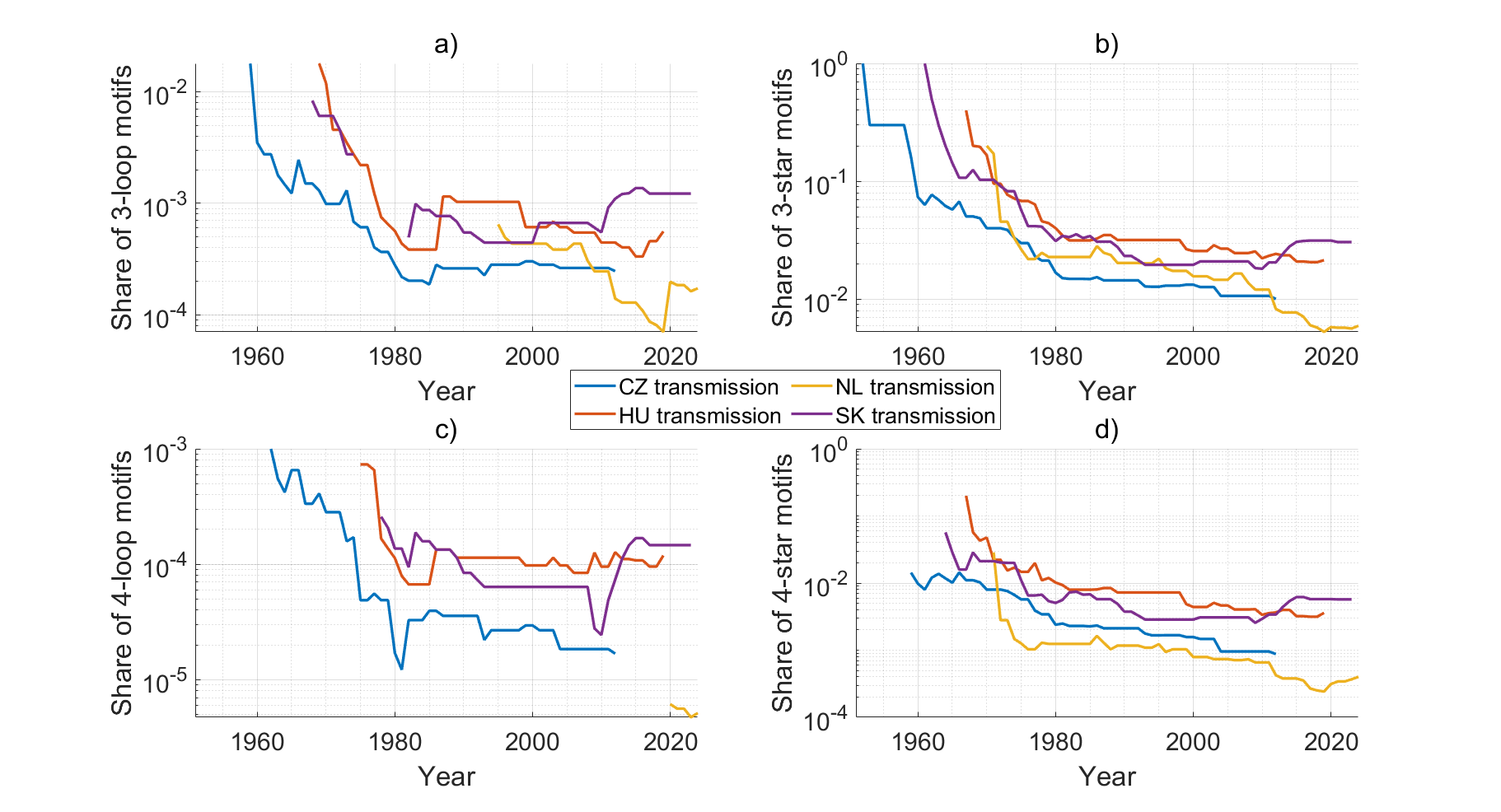}
     \caption{Share of various motifs in the four transmission systems.}\label{fig:Motifs}
 \end{figure}

Figure~\ref{fig:Motifs} presents the evolution of loop and star motif proportions over time. In general, we observe a relatively rapid decrease of loops and stars from 1950 to 1980 and eventually it stabilizes or decreases more slowly in recent years. The decreasing share of loops suggests that the networks are becoming more tree-like or hierarchical in structure. This results in the reduction of network redundancy and resilience. Although the network has become more simplified and cost-efficient, the tradeoff is its overall robustness. On the other hand, the reduction of star-like structures in the network implies the networks underwent decentralization and have become more uniformly connected and distributed. This results in a less efficient network (as seen in Figures~\ref{fig:CNA_subtransmission}(i), ~\ref{fig:CNA_transmission_alt}(i)) as hubs afford shorter paths, but at the same time, the network robustness has increased.

For both types of motifs, their trends experience an initial decrease in the earlier years and later on stabilizing in recent years. The initial decline can represent a phase of network pruning and simplification. The second phase, where the share of motifs seem to be constant over the years represent a state of network equilibrium where there are a certain number of loops enough to preserve resilience and stars structures to facilitate efficiency and coordination. This may be considered an optimal state wherein the networks are neither overly redundant, centralized, or sparse. These structural changes reflected in the fraction of loops and stars affect the clustering and path lengths in the network which can also explain how much faster the networks become much more small-world (based on $\sigma$) and for the decreasing $\omega $ values indicating change from random-like (shorter paths) to more small-world (see Figure~\ref{fig:SW}).

\section{Discussion}\label{sec4}

\begin{figure}[H]
     \centering    \includegraphics[width=\columnwidth ]{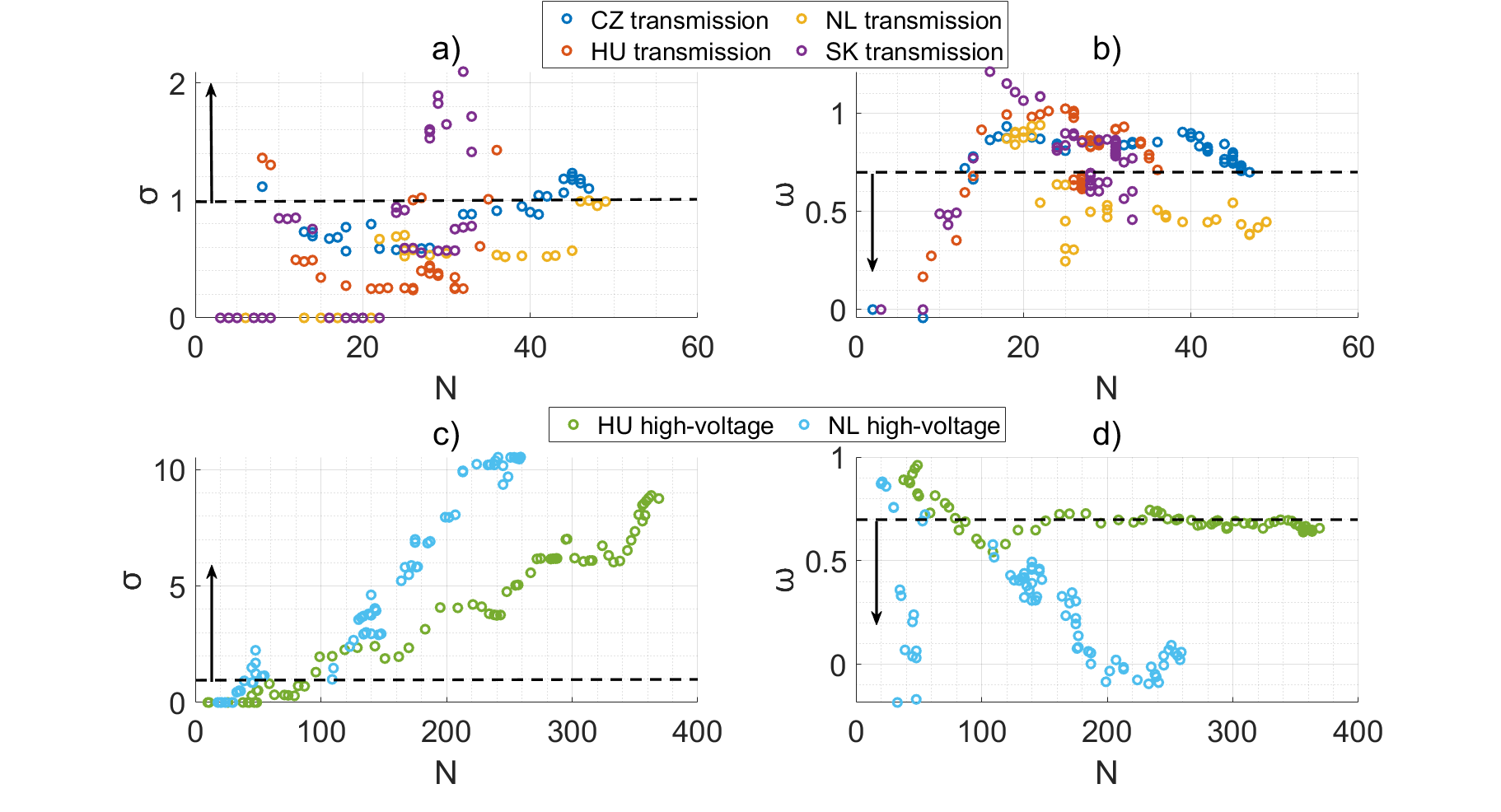}
     \caption{Small-worldness metrics as a function of node number}\label{fig:NvsSW}
 \end{figure}

In Figure~\ref{fig:NvsSW} we show the results of our analyses on the four datasets similar to what was done in a previous work~\cite{Hartmann2021}. Here, we slowly build the networks by adding nodes and lines of particular voltage values incrementally. Here, we can observe that as more elements are added, both the transmission and HV networks become more small-world with $\sigma \gtrsim 1$ and $\omega \lesssim 0.7$. In general, Figure~\ref{fig:NvsSW} presents the same behavior for the HV networks with $\sigma$ rising faster for both HU and NL. However, for $\omega$ we see that HU stabilizes while NL continues to drop. This may imply that for HU there is a balance of expansion and clustering of nodes. On the other hand, for NL the addition of nodes clustered together (high clustering, shorter paths) is faster than expanding the network coverage and thus, $\omega$ drops faster.

One may notice the differing behaviors of $\sigma$ and $\omega$ in Figures~\ref{fig:SW} and ~\ref{fig:NvsSW} which is more apparent in the case of the HV networks of HU and NL: $\sigma$ trends seem to rise very quickly as more nodes and lines are added. On the other hand, $\omega$ seems to be more stable. Take the case of HU, which fluctuates around $0.7$ consistently over the years. This inconsistency may be due to the fundamental differences in how the metrics are calculated. The metric $\sigma$ is sensitive to the changes in the clustering i.e., $C/C_r > L/L_r$ (Equation~\ref{eqn:sigma}). In contrast, $\omega$, being a ratio of normalized quantities, can remain stable, especially when clustering and path length change proportionally. Another reason is that $\sigma$ is highly sensitive to small increases in clustering especially when $C_r$ is very small (as in the case of sparse networks), whereas for $\omega$ the clustering is being compared to that of a lattice (Equation~\ref{eqn:omega}) where the clustering coefficient is high, making it less responsive to the same change. Having said these, $\sigma$ and $\omega$ have differing baselines: $\sigma$ compares both clustering and path to a random graph, while $\omega$ uses a lattice baseline for clustering and a random one for path. Therefore, proportional changes in network structure do not affect the two metrics in the same way.

\begin{figure}[H]
     \centering    \includegraphics[width=\columnwidth ]{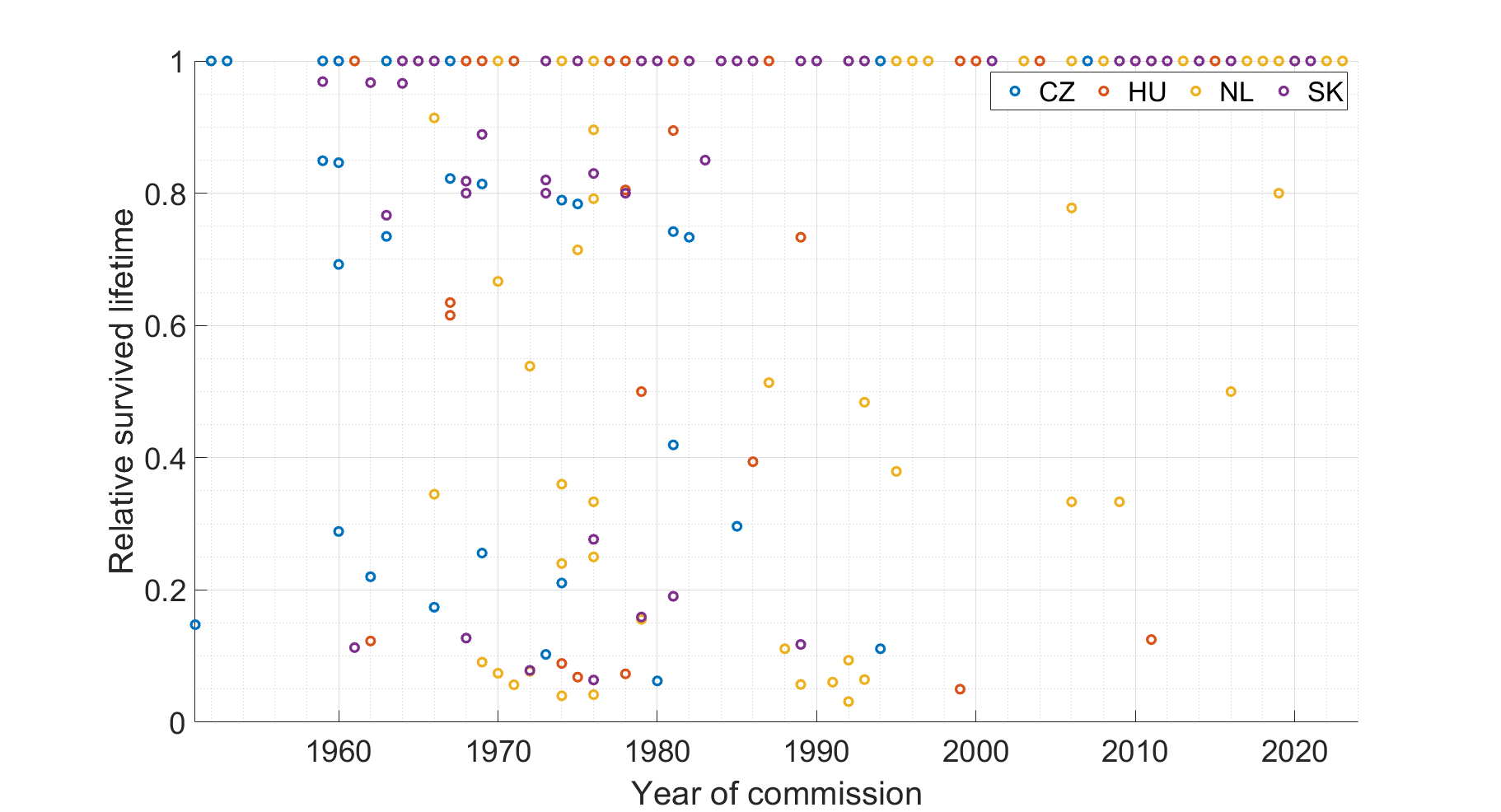}
     \caption{Relative survived lifetime of transmission lines}\label{fig:Survival}
 \end{figure}

To better understand the longevity of topological formations, we plotted the relative survived lifetime of each transmission line against their year of commissioning in Figure~\ref{fig:Survival}. It can be seen that there is a group of "under-performing" topological formations, whose lifetime is below approx. 20\% of their expected lifetime. (This translates as the maximal expected lifetime is the difference between the last year of the corresponding dataset and the year of commissioning.) Two separate group of datapoints can be identified. Between 1960-1980, all grids have a significant amount of "underperformers", while the NL grid also shows such elements in the 1990s.

As it was discussed in Section\ref{sec3.2}, the first period refers to the rapid expansion of transmission grids. In this era, the primary aim was to establish long-range interconnections, which in some cases led to lines of hundreds of kilometers. Fast forward a decade, and the increasing power needs necessitated more connections between transmission and sub-transmission levels, leading to the splitting of these connections. Also, there are numerous examples of lines commissioned on 220 kV or lower nominal voltages and upgraded to 400 kV in a few years.

\begin{figure}[H]
     \centering    \includegraphics[width=\columnwidth ]{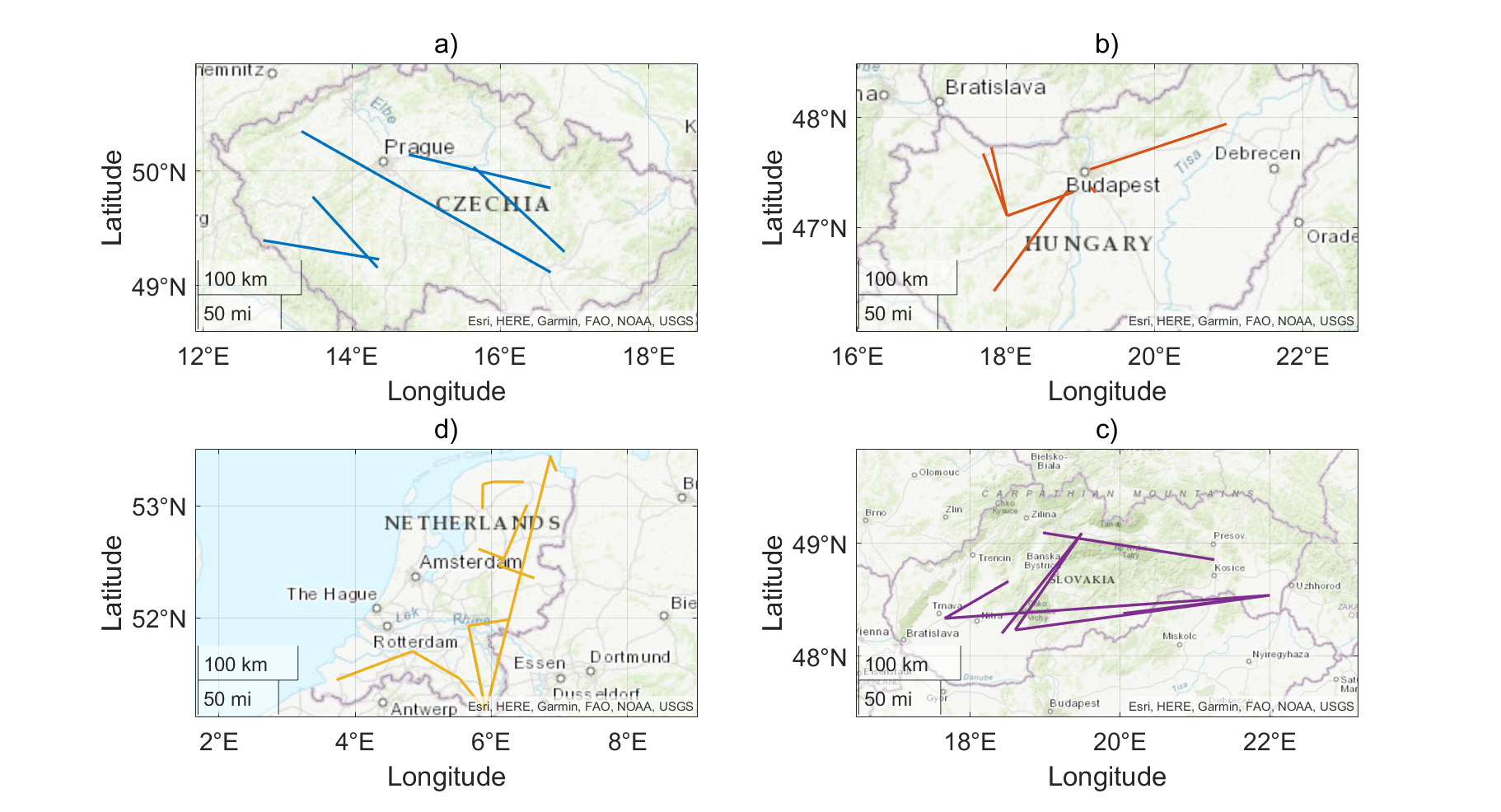}
     \caption{Topological elements changed in the first 20\% of their possible maximum lifetime}\label{fig:Abandoned}
 \end{figure}

Figure~\ref{fig:Abandoned} showcases these transmission lines in the four countries, also confirming that many of them were connections spanning over large distances.

\begin{figure}[H]
     \centering    \includegraphics[width=\columnwidth ]{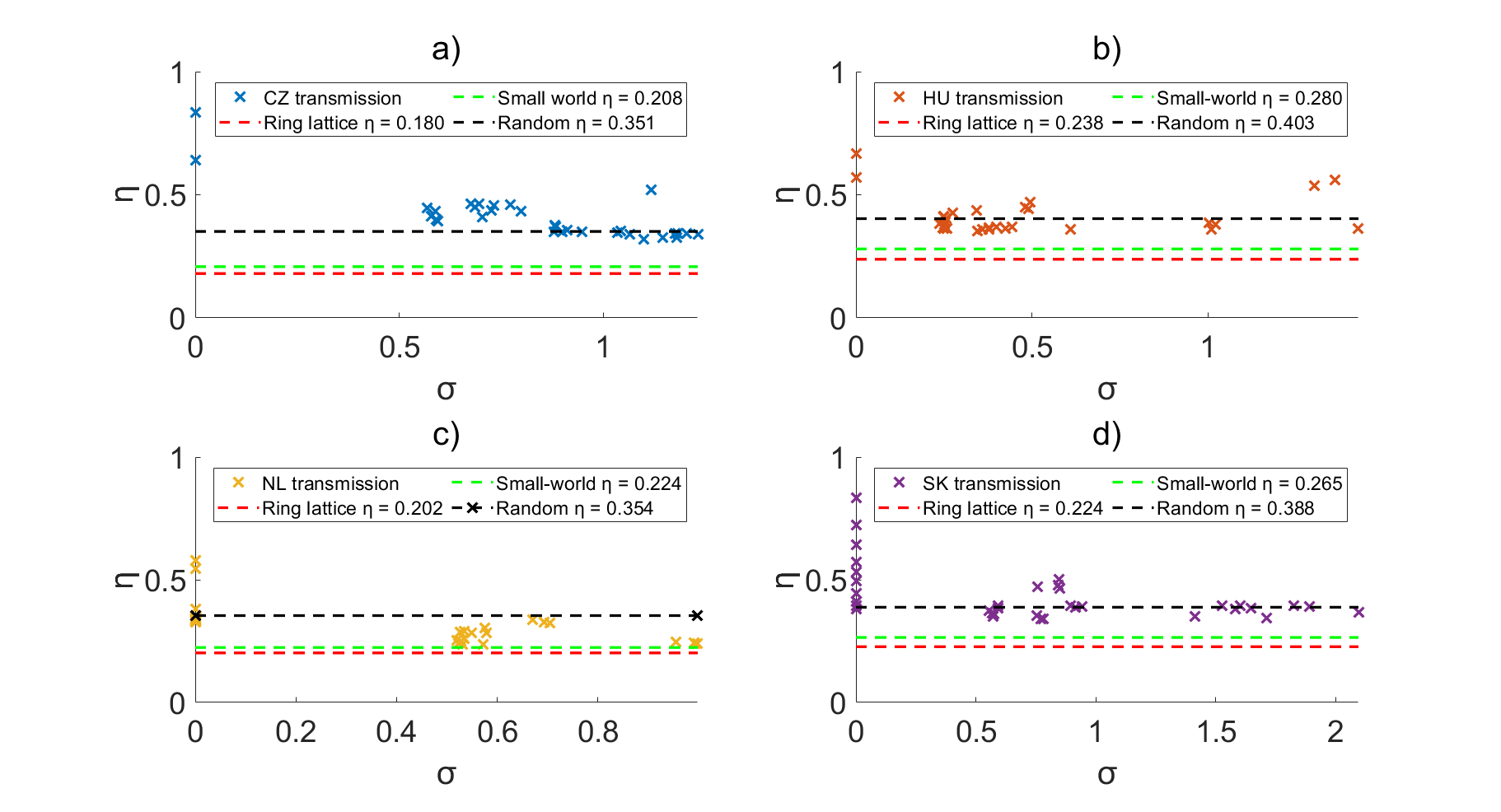}
     \caption{Relationship between the transmission ($w \geq 220$ kV) networks' smallworldness $\sigma$ to the global efficiency $\eta$.}\label{fig: SigmavsEff}
 \end{figure}

 In Figure~\ref{fig: SigmavsEff}, we compare the efficiency of the empirical transmission networks ($w \geq 220$ kV) to that of some representative theoretical networks: random, small-world, and ring lattice (this is what is used in smallworld studies) with $N$ and $E$ equal to the average number of nodes and edges of the empirical networks over the years. The random network can be expected to have the highest global efficiency because its connections result in very short average path lengths leading to high inverse distance and thus high efficiency. On the other hand, small world networks contain a few long-ranged connections in addition to the regular lattice, thereby maintaining high clustering while reducing path lengths (although lower than that of a random network). Finally, we have the ring lattice which we cannot expect to have high efficiency as nodes are only connected to their immediate neighbors in a ring. This results to very long path lengths between distant nodes and thus the lowest global efficiency among the reference networks. Here, we see that the empirical transmission networks span the region between smallworld and random, and in some cases, some years (one datapoint represents a year), the empirical have higher global efficiency than that of a random network. This is because of the fact that, power networks, are after all purposefully structured and not just randomly connected. Random networks lack geographic and engineering logic which means they have poor redundancy. The more efficient real-world networks reflect the optimization in minimizing distance and ensuring the reliability, balancing long-range power delivery and local robustness.

\section{Conclusions}

In this paper we examined a harmonized dataset of transmission grid evolution as observed in four European countries. Our multi-decade dataset with yearly resolution created a unique opportunity to perform a cross-system meta-analysis of the evolving networks. We computed complex network metrics, searched for simple network motifs, and did a temporal analysis, to reveal how structural features of these network evolve. We also linked aging and asset renewal to these topological motifs.

Our results empirically prove that transmission and sub-transmission infrastructure shows a remarkably similar structure in the analyzed countries, regardless of their significant differences in size, population, topography and history. We presented, that the networks were expanding, and their complexity was also increasing. The selected panel of complex network metrics shows a stabilization after the first 25-30 years of transmission grid evolution.

These findings underline that the power infrastructure is deliberately engineered, following various optimization targets, while balancing between the thin line of creating robust and flexible, yet still efficient topologies. We also coupled the topological characteristics with evidence from longitudinal data to provide an insight to topology control strategies on the long-run. Our results bridge an important gap in understanding the part of the grid in providing flexibility for future power systems.

Future work of the authors will set the focus on how temporal-network approaches can be applied to this comprehensive dataset, studying time-respecting paths and topological persistence.

By coupling evidence from longitudinal data with topological characteristics, the study contributes to the understanding of topology control strategies in empirical evolution patterns.

\label{sec5}

\section{Acknowledgement}\label{sec6}
Bálint Hartmann acknowledges the support of the Bolyai János Research Scholarship of the Hungarian Academy of Sciences (BO/131/23), and the National Research Excellence Programme of the National Research, Development and Innovation Office (ADVANCED 150405).

The authors would like to thank the following people for their contribution to assembling the historical grid database: Rens Baardman, Attila Dervarics, Zoltán Feleki, József Hiezl, János Nemes, Imre Orlay, János Rejtő, Viktória Sugár.





\bibliographystyle{elsarticle-num}
\bibliography{references.bib}

\end{document}